\documentclass[twocolumn,prb,amsmath,amssymb,floatfix,superscriptaddress]{revtex4-2} 

\usepackage{mathtools}
\usepackage[english]{babel}
\usepackage{amsfonts}
\usepackage{amsmath}
\usepackage{tikz}
\usepackage{xcolor}
\usepackage[colorlinks, citecolor={blue}, urlcolor={blue}, linkcolor={red!80!black}]{hyperref}
\usepackage{orcidlink}
\usepackage{tabularray}

\usepackage{siunitx}
\usepackage{physics}
\ExplSyntaxOn
\msg_redirect_name:nnn{siunitx}{physics-pkg}{none} 
\ExplSyntaxOff

\begin{document}

\title{Superconducting properties of transition metal dichalcogenides in proximity to a conventional superconductor}

\author{{Florian Kayatz}\,\orcidlink{0009-0005-3572-3561}}
\email{florian.kayatz@physics.uu.se}
\affiliation{Department of Physics and Astronomy, Uppsala University, Box 524, SE-751 20 Uppsala, Sweden}
\affiliation{Wallenberg Initiative Materials Science for Sustainability, Uppsala University, SE-751 20 Uppsala, Sweden}

\author{{Annica M. Black-Schaffer}\,\orcidlink{0000-0002-4726-5247}}
\affiliation{Department of Physics and Astronomy, Uppsala University, Box 524, SE-751 20 Uppsala, Sweden}
\affiliation{Wallenberg Initiative Materials Science for Sustainability, Uppsala University, SE-751 20 Uppsala, Sweden}

\author{{Jorge Cayao}\,\orcidlink{0000-0001-6037-6243}}
\affiliation{Department of Physics and Astronomy, Uppsala University, Box 524, SE-751 20 Uppsala, Sweden}

\date{\today}

\begin{abstract}
    Transition metal dichalcogenides (TMDs) hold relevance for spin-triplet superconducting phases due to their inherent Ising spin-orbit coupling, but the majority of studies have so far focused on oversimplified models.
    In this work, we consider a TMD monolayer using a three-orbital model with anisotropic couplings and investigate the emergent superconducting properties when it is placed in proximity to a conventional spin-singlet $s$-wave superconductor.
    We find that the multiorbital nature of the TMDs lead to superconducting gaps not only at zero energy but also at higher energies, so-called hybridization gaps, which exhibit a complex structure due to the anisotropic couplings, challenging their spectral measurement.
    Moreover, we find that the inherent Ising spin-orbit coupling induces a spin splitting and a spin polarization along the $z$ direction, which correlates with the emergence of mixed spin-triplet superconducting pairs.
    These spin-triplet pair correlations appear in the monolayer as a proximity-induced effect due to the impact of the Ising spin-orbit field on conventional spin-singlet $s$-wave superconductivity.
    Taking realistic parameters for a $\text{MoS}_2$ monolayer, we show that the Ising field is strong enough to induce spin-triplet pair correlations of the same magnitude as their spin-singlet counterparts.
    We also include Rashba spin-orbit coupling, naturally emerging in a heterostructure and find that it induces equal spin-triplet superconducting pairs that compete with the mixed spin-triplet pairs induced by the Ising spin-orbit coupling.
    Our findings help understand the superconducting properties of TMDs in proximity to conventional superconductors.
\end{abstract}

\maketitle

\section{\label{sec:introduction}Introduction}

The progress made since the discovery of graphene \cite{novoselov_graphene_2004, novoselov_graphene_2005, zhang_graphene_2005} has opened up the possibility of exploring quasi two-dimensional properties of layered materials down to single layers.
One interesting example is the family of transition metal dichalcogenides (TMDs), where monolayers are now routinely fabricated \cite{novoselov_two_dimensional_2005}.
TMDs have been the focus of extensive research because of their unique properties, making them promising candidates for various applications, such as in nanoscale electronics \cite{wand_tmdcs_2012}, optoelectronics \cite{wand_tmdcs_2012}, valleytronics \cite{schaibley2016}, and spintronics \cite{ahn2020}.

In general, TMD monolayers exhibit properties that are distinct from their bulk versions, including a transition from an indirect to a direct band gap with reduced layer count for semiconducting TMDs \cite{wand_tmdcs_2012}.
Even more interesting, and unlike their bulk counterparts, the broken inversion symmetry leads to a strong antisymmetric spin-orbit coupling in monolayers, mainly originating from the $d$ orbitals of the transition metal atoms, known as Ising spin-orbit coupling (SOC) \cite{zhu_giant_soc_2011,frigeri_2004, zhou_spin-orbit_2019}.
The corresponding Ising field has opposite signs in the $K$ and $K'$ valleys in the Brillouin zone, thus preserving time-reversal symmetry \cite{tang_magnetic_2021} and leading to valley Hall effects \cite{xiao_coupled_2012}.

Beyond normal state phases, TMD monolayers have also been studied in the superconducting realm.
For instance, both $\text{NbSe}_2$ and $\text{TaS}_2$ have been experimentally verified to host intrinsic superconductivity down to single layers \cite{frindt_superconductivity_1972, xi_cdw_nbse2_2015, xi_ising_nbse2_2016, barrera_tuning_2018}, while $\text{MoS}_2$ can become superconducting under ionic gating \cite{lu_mos2_sc_2015}.
In these monolayers, the large Ising SOC polarizes the spins out of plane, supporting the presence of unconventional Ising pairing \cite{xi_ising_nbse2_2016}, which results in high critical fields \cite{xi_ising_nbse2_2016, barrera_tuning_2018, lu_mos2_sc_2015, saito_sc_gated_mos2_2016} well above the Pauli paramagnetic limit \cite{clogston_upper_1962, chandrasekhar_pauli_1962}.
Such unconventional pairing states are generally known to exist in non centrosymmetric superconductors, where the inherent absence of inversion symmetry leads them to host, for example, spin-triplet correlations \cite{Smidman_2017, Bauer_Sigrist_2012}.
Specifically for TMD monolayers, this has led to a plethora of studies investigating intrinsic superconductivity \cite{PhysRevB.101.014510, PhysRevX.10.041003,PhysRevB.104.L060501, 10.1063/5.0153345, Siegl2025, PhysRevResearch.7.013290}.
These studies have shown through symmetry analysis of the gap function that the presence of Ising SOC and inversion symmetry breaking results in the emergence of mixed spin-triplet superconducting correlations in TMD monolayers \cite{PhysRevB.98.144518,PhysRevB.101.224503}, as well as that the possibility of equal spin-triplet correlations is necessitated on the presence of Rashba SOC or magnetic fields \cite{PhysRevB.99.180505, PhysRevB.101.224503}.
However, the exact pairing mechanism has not yet been conclusively determined \cite{Hsu2017, Das2023, 6rc6-mm3b}.

Superconductivity can be induced in an even wider range of TMD monolayers, including those that are not intrinsically superconducting, via the superconducting proximity effect when placed in contact with a superconductor \cite{holm_proximity_1932, budzin_proximity_2005, RevModPhys.77.1321, tanaka2011symmetry, Cayao2020odd, PhysRevMaterials.8.104801}, as already reported for $\text{MoS}_2$ \cite{trainer_proximity_sc_mos2_2020}.
Building upon the well-established theoretical description of the conventional superconductors with which the TMD monolayer is in contact with, this allows for a more controlled way to exploit the potential of TMDs for realizing exotic superconducting phases \cite{PhysRevB.90.100505, PhysRevB.89.155317,wakatsuki_proximity_2016}.
Since for a minimal model of proximity-induced superconductivity in a TMD monolayer the underlying symmetry group is the same as for the intrinsic case, the conclusions drawn from the classification of the gap function in intrinsic superconductivity \cite{PhysRevB.98.144518, PhysRevB.99.180505,  PhysRevB.101.224503, Siegl2025, 6rc6-mm3b} are also transferable to the proximity-induced case.
Therefore, also proximity to a conventional spin-singlet $s$-wave superconductor is expected to offer unprecedented opportunities because TMDs are both multiorbital systems and possess intrinsic Ising SOC, which are expected to favor both spin-triplet \cite{gorkov_superconducting_2001, PhysRevLett.103.020401, Bauer_Sigrist_2012, reeg_proximity-induced_2015, PhysRevB.93.180501,  Smidman_2017, PhysRevB.98.075425,  PhysRevB.102.214513} and odd-frequency superconducting pairing \cite{RevModPhys.77.1321,tanaka2011symmetry,linder_odd-frequency_2019,triola_role_2020,Cayao2020odd,tanaka2024,fukayaCayaoReview2025_IOP}.
This suggests that proximitized monolayer TMDs are a promising platform for the realization of spin-triplet superconductivity as an alternative to the more commonly used superconductor-ferromagnet interfaces \cite{budzin_proximity_2005, singh_spin_valve_2015}.
While these ideas have motivated some studies, previous work on proximity-induced superconductivity in TMDs has focused on low-energy $\mathbf{k}\cdot \mathbf{p}$ models that approximate the bands only around the $K$ and $K'$ valleys.
Since such models are only approximate around these valleys, it is natural to wonder about the implications of a more realistic description of TMD monolayers when proximitized with a conventional superconductor.

In this work, we consider a TMD monolayer proximitized with a spin-singlet $s$-wave superconductor (SC) in a heterostructure, see Fig.~\ref{fig:model}(a), and investigate the emergent superconducting properties.
In particular, we model the TMD monolayer based on a three-orbital model having both anisotropic couplings and Ising SOC, which reproduces well the band structure of TMD monolayers obtained from first-principles calculations \cite{liu_three-band_2013} and allows us to characterize the superconducting correlations over the whole Brillouin zone.
We first demonstrate that the heterostructure develops a two-gap structure around zero, including a larger gap due to the native SC and a smaller one induced in the TMD.
Additionally, the multiorbital structure of TMD monolayers gives rise to gaps due to hybridization at band crossings between electron (hole) superconducting and hole (electron) TMD bands at higher energies, which are distinguishable in the spectral function.
In contrast to the naively expected clean hybridization gaps in simpler multiband models \cite{komendova_odd_frequency_2015,triola_general_2016,komendova_odd-frequency_2017,triola_role_2020,PhysRevB.101.214507,PhysRevB.103.104505,tang_magnetic_2021,PhysRevB.108.214510,PhysRevB.109.205406,fu2025light,fu2025floquet, PhysRevB.108.214510}, we show that more realistic modeling makes these hybridization gaps rather difficult to identify in the density of states due to the anisotropic couplings between orbitals.
Nevertheless, the opening of hybridization gaps is a result of proximity-induced superconductivity in the TMD.

Next, to unveil the nature of the induced superconductivity, we carry out a full symmetry classification of the superconducting correlations and identify that the combined action of the multiorbital nature and the Ising SOC enables the formation of eight different classes of proximity-induced pairs.
In particular, we find the emergence in the TMD of mixed spin-triplet superconducting correlations in agreement with previous studies \cite{PhysRevB.98.144518, PhysRevB.99.180505, PhysRevB.100.054503, PhysRevB.101.224503, Siegl2025, 6rc6-mm3b}, which are present as long as there exists a finite spin-polarization along $z$ due to the Ising SOC.
These induced mixed spin-triplet pair correlations are often comparable in magnitude to the spin-singlet pair components and can even become larger than the spin-singlet pairing, especially in the energy regions between spin-split hybridization gaps.
To account for possible breaking of the out-of-plane mirror symmetry in proximitized TMD monolayers \cite{lu_janus_2017, lee_rashba_mos2_2015, chen_probing_2023}, we also include Rashba SOC and show the appearance of equal spin-triplet superconducting pairs, followed by a discussion of the competing influences when determining the strength of the two types of spin-triplet pairings when both Ising and Rashba SOC are present in the system.

The remainder of this work is organized as follows.
In Sec.~\ref{sec:model} we present an accurate three-orbital model of a TMD monolayer in proximity to a conventional superconductor.
Here we also discuss how we obtain the spectral signatures and superconducting pair amplitudes.
In Sec.~\ref{sec:results_ising} we investigate the signatures of the Ising SOC in both the spectral function and the density of states, while in Sec.~\ref{sec:results_pair_correlations} we present the induced superconducting correlations.
In Sec.~\ref{sec:results_rashba}, we study the additional effect of Rashba SOC.
Finally, in Sec.~\ref{sec:conclusion} we present our conclusions. For completeness, Appendix~\ref{sec:tmdmodel} shows the TMD model details, Appendix~\ref{sec:scgap} discusses minimal models for the induced  superconducting gaps, Appendix~\ref{sec:appendix_t} assesses the effect of the tunneling between TMD and SC, Appendix~\ref{sec:appendix_rashba} shows equal spin-triplet pairs under Rashba SOC, and Appendix~\ref{sec:appendix_sc} analyzes the formation of pair amplitudes in the SC.

\section{\label{sec:model} Model and Methods}

We here present the model of a TMD monolayer in proximity to a conventional spin-singlet $s$-wave superconductor and then discuss the methodology for obtaining the spectral signatures and superconducting correlations.

\subsection{Model Hamiltonians}
\begin{figure}
    \centering
    \includegraphics[width=\columnwidth]{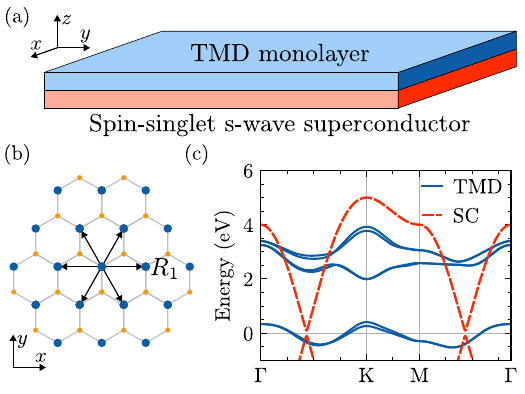}
    \caption{(a) Schematic picture of a monolayer of a TMD proximitized by a spin-singlet $s$-wave SC in a heterostructure. (b) Top-down view of the hexagonal TMD lattice structure with large blue transition metal and small orange chalcogen atoms. The TMD is modeled using a three-band tight-binding model including the $d_{z^2}$, $d_{xy}$ and $d_{x^2-y^2}$ orbitals of the blue transition metal atoms, where the black arrows and $R_1$ indicate the nearest-neighbor atoms. (c) Band structure for the monolayer TMD $\text{MoS}_2$ and for a conventional $s$-wave superconductor on a triangular lattice (electron bands shown for TMD, both electron and hole bands shown for the superconductor).}
    \label{fig:model}
\end{figure}

We consider a TMD monolayer coupled to a conventional superconductor, as sketched in Fig.~\ref{fig:model}(a). This heterostructure is modeled by the Hamiltonian
\begin{equation}
    H = H_{\text{TMD}} + H_{\text{SC}} + H_{T}\,, \label{eq:htmdsc}
\end{equation}
where $H_{\text{TMD}}$ and $H_{\text{SC}}$ describe the TMD monolayer and SC, respectively, while $H_{T}$ models the tunneling coupling between the SC and TMD.
All elements are treated within a tight-binding description, as explained below.

\subsubsection{TMD monolayer}
The TMD monolayer exhibits a hexagonal lattice structure with a two-atom basis, as shown in the top-down view of the TMD monolayer in Fig.~\ref{fig:model}(b).
We describe the monolayer with a three-band tight-binding model that reproduces the band structure obtained from first principles \cite{liu_three-band_2013} and includes the three orbitals $d_{z^2}, d_{xy}$, and  $d_{x^2-y^2}$ of the transition metal atoms (blue).
To simplify our notation, we denote these orbitals with integers $a = 1,2,3$.
An analysis of the crystal symmetries shows that the orbital $d_{z^2}$ transforms under the trivial representation $A^{'}_1$, while the orbitals $d_{xy}$ and $d_{x^2-y^2}$ transform under the two-dimensional irreducible representation $E^{'}$, respectively, of the $\text{D}_{3\text{h}}$ point group of the lattice \cite{mattheiss_band_1973,liu_three-band_2013}.
In the basis $\text{spin} \otimes \text{orbital}$, the TMD Hamiltonian $H_{\text{TMD}}$ in Eq.~\eqref{eq:htmdsc} is given by
\begin{equation}
    H_{\text{TMD}} = H_{\text{TMD}}^{(0)} + H_{\text{Ising}} + H_{\text{Rashba}}\,,
\end{equation}
where $H_{\text{TMD}}^{(0)}$ describes the kinetic term, $H_{\text{Ising}}$ is the Ising SOC, and $H_{\text{Rashba}}$ the Rashba SOC.
To be more precise, $H_{\text{TMD}}^{(0)}$ up to third-nearest neighbor hoppings is given by \cite{liu_three-band_2013}
\begin{equation}
    H_{\text{TMD}}^{(0)} = \sigma_0 \otimes \bigg( \sum_{a, b, \mathbf{k}} E_{\mathbf{k}}^{ab} c^\dagger_{a, \mathbf{k}} c_{b, \mathbf{k}} - \mu \bigg), \label{eq:ham_tmd}
\end{equation}
where $c^{\dagger}_{a, \mathbf{k}} (c_{b, \mathbf{k}}) $ creates (annihilates) an electron with momentum $\mathbf{k}$ in orbital $a (b)$, and $\mu$ is the chemical potential measuring the bands' filling.
The exact form of the coefficients $E_{\mathbf{k}}^{ab}$ is detailed in Appendix~\ref{sec:tmdmodel}, including the numerical constants used to specifically model monolayer $\text{MoS}_2$.

The generic form of the intrinsic on-site Ising SOC $H_{\text{Ising}}$ is given by \cite{liu_three-band_2013, huang_generic_2024}
\begin{equation}
    H_{\text{Ising}} = \lambda \,\mathbf{S} \cdot \mathbf{L},
\end{equation}
where $\lambda$ is the strength of Ising SOC and $\mathbf{S} (\mathbf{L})$ is the spin(angular)-momentum operator. Since the three $d$ orbitals we are considering have ${m_z=-2,0,2}$ as their eigenvalues of the $L_z$ operator, only the $z$ component of the angular-momentum operator is nonzero \cite{liu_three-band_2013}.
Using the explicit form of the $d$ orbitals, we find
\begin{equation}
    H_{\text{Ising}} = \lambda \, S_z \otimes L_z = \lambda \, S_z \otimes \begin{pmatrix} 0 & 0 & 0 \\ 0 & 0 & 2i \\ 0 & -2i & 0 \end{pmatrix}. \label{eq:ham_ising}
\end{equation}
Considering both the kinetic term and the Ising SOC results in the band structure shown by solid blue lines in Fig.~\ref{fig:model}(c).

Furthermore, the interface in the $z$ direction breaks the out-of-plane mirror symmetry, which induces Rashba SOC in each of the orbitals of the TMD \cite{bychov_rashba_soc_1984, lee_rashba_mos2_2015}.
The real-space Hamiltonian describing this SOC for an orbital $l$ is given by \cite{huang_generic_2024}
\begin{equation}
    H_{\mathrm{Rashba}}(\mathbf{r}) = i \gamma_l \sum_{m,n} \hat{c}^\dagger_{l,m} \left(\mathbf{S} \times \mathbf{d}_{mn} \right)_z \hat{c}_{l,n},
\end{equation}
with strength $\gamma_l$ and the nearest-neighbor lattice vectors $\mathbf{d}_{mn}$ in units of the lattice constant $a$.
Here, the creation operator $\hat{c}_{l,m}^\dagger$ creates an electron in the orbital $l$ at site $m$.
The elements are explicitly given by
\begin{equation}
    \left(\mathbf{S} \times \mathbf{d}_{mn} \right)_z = \frac{1}{2}\left( \sigma_x \mathbf{d}_{mn,y} - \sigma_y \mathbf{d}_{mn,x} \right),
\end{equation}
with the spin Pauli matrices defined by $\sigma_z \ket{\pm} = \pm \ket{\pm}$.
The Rashba SOC strength can be estimated from conduction band measurements around the $K$ points \cite{zhou_spin-orbit_2019,lu_mos2_sc_2015} and we assume it to be constant over the Brillouin zone and for both irreducible representations $A^{'}_1$ and $E^{'}$.
Thus, we introduce the Rashba SOC strength $\gamma$ and transforming into reciprocal space, we arrive at
\begin{equation}
    H_{\text{Rashba}} = \left(f_x(\mathbf{k}) \sigma_y - f_y(\mathbf{k}) \sigma_x\right) \otimes \left(\gamma \mathbf{1}_3\right), \label{eq:ham_rashba}
\end{equation}
with the momentum-dependent coefficients given by \cite{zhou_spin-orbit_2019}
\begin{align}
    f_x(\mathbf{k}) & = \sin(2\alpha) + \sin(\alpha) \cos(\beta), \\
    f_y(\mathbf{k}) & = \sqrt{3} \sin(\beta) \cos(\alpha),
\end{align}
where ${\alpha =  k_x a/2}$ and ${\beta=\sqrt{3} k_y a /2}$.

\subsubsection{Superconductor and its coupling to the TMD}
We consider the superconductor to be a single-band spin-singlet $s$-wave superconductor.
To keep our notation unified, we label this additional band with the index $a = 4$.
The Hamiltonian of the superconductor is expressed as the sum of the normal state part $H_{\text{SC}}$ and the superconducting pairing term $\hat{\Delta}$, given by
\begin{align}
    H_{\text{SC}} & = \sum_{\mathbf{k},\sigma} \xi_\mathbf{k} c^\dagger_{4,\mathbf{k},\sigma} c_{4,\mathbf{k},\sigma}, \\
    \hat{\Delta}  & = \sum_{\mathbf{k}} \Delta c_{4,-\mathbf{k},\uparrow} c_{4,\mathbf{k},\downarrow} + \mathrm{H.c.},
\end{align}
where $c^\dagger_{4,\mathbf{k},\sigma}$ creates an electron with momentum $\mathbf{k}$ and spin $\sigma$ in the superconductor and $\Delta$ is the superconducting order parameter.
Moreover, $\xi_\mathbf{k}$ is the kinetic superconducting term, which we obtain by assuming for simplicity a triangular lattice with matching lattice constant $a$ as the TMD, given by
\begin{equation}
    \xi_{\mathbf{k}} = 2 t_{\text{SC}} \left(\cos 2\alpha + 2\cos \alpha \cos \beta\right) - \mu_{\text{SC}},
\end{equation}
with $t_{\text{SC}}$ being the nearest-neighbor hopping in the superconductor, and $\mu_{\text{SC}}$ the chemical potential.
This results in the superconducting band structure shown by red dashed lines in Fig.~\ref{fig:model}(c), which is symmetric around zero energy due to particle-hole symmetry.

Last, in order to induce superconductivity in the TMD via the proximity effect, we allow the electrons to tunnel between both systems, modeled by the tunneling Hamiltonian
\begin{equation}
    H_{T} = \sum_{\mathbf{k},\sigma} t c^{\dagger}_{1,\mathbf{k},\sigma} c_{4,\mathbf{k},\sigma} + \text{H.c.},
\end{equation}
with tunneling strength $t$.
In order to not break the inherent $\hat{C}_3$ symmetry of the crystal lattice underlying the model and because the orbital overlap between the $s$-wave orbital of the superconductor and the $d_{xy}$ and $d_{x^2-y^2}$ orbitals of the TMD is zero due to symmetry, we only consider tunneling from the band of the superconductor to the $d_{z^2}$ orbital in the TMD monolayer.

\subsubsection{Bogoliubov--de Gennes Hamiltonian}
To solve the full Hamiltonian, we express it within the Bogoliubov-de Gennes (BdG) formalism.
We start by rewriting the Hamiltonian given by Eq.~\eqref{eq:htmdsc} in Nambu space as
\begin{equation}
    \mathcal{H}_\text{BdG} = \frac{1}{2}\sum_{\mathbf{k}} \Psi^\dagger_{\mathbf{k}} H_\text{BdG}(\mathbf{k}) \Psi_{\mathbf{k}}\,,
\end{equation}
where ${\Psi^\dagger_\mathbf{k} = (\Phi^\dagger_\mathbf{k}, \Phi_{-\mathbf{k}}^T)}$, with $\Phi^\dagger_\mathbf{k}$ forming the basis of our normal-state Hamiltonian, and
\begin{equation}
    H_\text{BdG}(\mathbf{k}) = \begin{pmatrix}
        H(\mathbf{k})        & \hat{\Delta}      \\
        \hat{\Delta}^\dagger & -H^*(-\mathbf{k})
    \end{pmatrix}\,, \label{eq:ham_bdg}
\end{equation}
being the BdG Hamiltonian.
This Hamiltonian is the central object for investigating the spectral signatures and superconducting correlations.

For the numerical calculations, we consider parameters for the TMD monolayer consistent with realistic modeling \cite{liu_three-band_2013}, and unless otherwise stated we choose $t = 0.2\,\mathrm{eV}$, $\lambda = 0.073\,\mathrm{eV}$, $a = 3.19\,\text{\AA}$, $\mu = -0.4\,\mathrm{eV}$. The additional parameters needed to model monolayer $\text{MoS}_2$ are presented in Appendix~\ref{sec:tmdmodel}. We further discuss the influence of the tunneling strength $t$ on the results in Appendix~\ref{sec:appendix_t}.
For the superconductor, we consider $t_{\text{SC}} = 1\,\mathrm{eV}$, $\mu_{\text{SC}} = 2\,\mathrm{eV}$, and $\Delta = 0.1\,\mathrm{eV}$ for a large band width and generic filling. We choose a large value for the superconducting gap $\Delta$ to make the numerical computations feasible, but which does not affect our main results.

\subsection{Superconducting correlations \label{sec:methods_pair_correlations}}
To characterize the emergent superconducting correlations, we use the anomalous electron-hole Green's functions \cite{zagoskin,mahan2013many} for the BdG Hamiltonian in Eq.~\eqref{eq:ham_bdg}.
We calculate the total Green's function as
\begin{equation}
    \label{eq:GF}
    \begin{split}
        G(\mathbf{k},z) & = {\left(z - H_{\text{BdG}}(\mathbf{k})\right)}^{-1} \\
                        & = \begin{pmatrix}
                                G_0(\mathbf{k},z)     & F(\mathbf{k},z)         \\
                                \bar{F}(\mathbf{k},z) & \bar{G}_0(\mathbf{k},z)
                            \end{pmatrix},
    \end{split}
\end{equation}
where $z$ represents complex frequencies, $G_0 (\bar{G}_0)$ is the normal electron-electron (hole-hole) Green's function, and $F (\bar{F})$ is the anomalous electron-hole (hole-electron) Green's function.
The normal component allows us to calculate the spectral function and density of states (see below), while the anomalous Green's function gives us the superconducting correlations.
Since Eq.~\eqref{eq:GF} includes spin, orbital, and Nambu subspaces, both normal and anomalous parts are matrices.
In particular, the elements of the anomalous Green's functions $F$ can be denoted as \cite{zagoskin}
\begin{equation}
    F_{ab}^{\sigma \sigma'}(\mathbf{k}, z) = \langle \hat{T} c_{a,\mathbf{k},\sigma}(z) c_{b,-\mathbf{k},\sigma'}(0) \rangle,
\end{equation}
where $\hat{T}$ is the time-ordering operator, $\sigma, \sigma'$ label the spin, and ${a,b \in\{1,2,3,4\}}$ label the orbital degree of freedom.
Here, $F_{ab}^{\sigma \sigma'}(\mathbf{k}, z)$ is often also called superconducting pair amplitude or just pair amplitude. These pair amplitudes have a clear physical meaning: They represent the wave functions of Cooper pairs and therefore offer direct access to their symmetries \cite{tanaka2011symmetry,linder_odd-frequency_2019,Cayao2020odd}.

As a consequence of Fermi-Dirac statistics, the superconducting correlations are antisymmetric under the exchange of all degrees of freedom \cite{RevModPhys.77.1321,tanaka2011symmetry,linder_odd-frequency_2019,triola_role_2020,Cayao2020odd,tanaka2024,fukayaCayaoReview2025_IOP}
\begin{equation}
    F_{ab}^{\sigma \sigma'}(\mathbf{k}, z) = -F_{ba}^{\sigma' \sigma}(-\mathbf{k}, -z),
\end{equation}
which involves spins, momentum, orbitals, and frequency.
This leads to the so-called SPOT classification \cite{linder_odd-frequency_2019}
\begin{equation}
    S \times P \times O \times T = -1\,, \label{eq:spot_rule}
\end{equation}
where $S$ exchanges the spin, $P$ the momentum, $O$ the orbital and $T$ the time (frequency) degrees of freedom between the two electrons constituting a Cooper pair or equivalently the superconducting pair amplitude.
Thus, the superconducting correlations are either symmetric $(+1)$ or antisymmetric $(-1)$ under the exchange of these individual degrees of freedom, but the product of all four symmetries has to be $-1$.
In this regard, conventional spin-singlet $s$-wave superconductivity is odd in spin ($S=-1$) and even in momentum ($P=+1$), orbital ($O=+1$), and time ($T=+1$).
However, the antisymmetry condition via the SPOT classification allows for many more possibilities, including so-called odd-frequency superconductivity \cite{RevModPhys.77.1321,tanaka2011symmetry,linder_odd-frequency_2019,triola_role_2020,Cayao2020odd,tanaka2024,fukayaCayaoReview2025_IOP}, which is odd in the relative time of the paired electrons ($T=-1$) or, equivalently, in frequency.
The pair amplitudes can also be parametrized for specified band indices $\mathbf{F}_{ab}(\mathbf{k}, z)$ in terms of Pauli matrices as
\begin{equation}
    \mathbf{F}_{ab}(\mathbf{k}, z) = \left(F^{(0)}_{ab}(\mathbf{k}, z) \sigma_0 + \mathbf{d}_{ab}(\mathbf{k}, z) \cdot \boldsymbol{\sigma}\right) i \sigma_y,
\end{equation}
where $F^{(0)}_{ab}$ describes the spin-singlet part, while the vector $\mathbf{d}_{ab} = (F^{(x)}_{ab}, F^{(y)}_{ab}, F^{(z)}_{ab})$ characterizes both the equal-spin ($x$ and $y$) and mixed spin-triplet ($z$) correlations.
Taking this into account, we find the degrees of freedom present in the TMD-SC heterostructure allow four symmetry classes of spin-singlet superconducting pair amplitudes and four of spin-triplet type.
All these pair symmetry classes are listed in Table~\ref{tab:pair_correlations_symmetries}.
Here, the presence of spin-triplet pair amplitudes is in agreement with previous works that have studied the classification of the gap function in TMD monolayers according to the corresponding irreducible representations, see for example Refs.~\cite{PhysRevB.98.144518, PhysRevB.99.180505, PhysRevB.100.054503, PhysRevB.101.224503, Siegl2025, 6rc6-mm3b}.
It is further possible to find superconducting pair amplitudes due to electron pairs only within the TMD (SC), which we refer to as local, but also pairs forming between electrons in the TMD and SC, which we refer to as nonlocal.
Because we consider a spin-singlet $s$-wave superconductor, a priori only the local spin-singlet part $F^{(0)}_{44}$ in the SC is nonzero.
Thus, all other pair amplitudes are emergent phenomena due to the interplay between Ising SOC, multiorbital nature of TMDs, and conventional superconductivity.

It is also worth noting that spin-triplet pair amplitudes might induce a nonzero Cooper pair polarization defined as \cite{sigrist_unconv_supercon_1991}
\begin{equation}
    P_{ab} = i (\mathbf{d}_{ab} \times \mathbf{d}_{ab}^*). \label{eq:spin_pol}
\end{equation}
Here, $P_{ab}$ becomes nonzero if time-reversal symmetry is broken \cite{he_spin_supercurrent_2019}, but systems with additional internal degrees of freedom, such as multiple orbitals, can also host a finite Cooper pair polarization.
Thus, Cooper pair polarization can be seen as a signal of emergent physics.
Since the TMD and SC form a heterostructure, superconducting correlations along with their Cooper pair polarization are expected to be induced into the TMD by the proximity effect, while the TMD is also expected to affect the SC by means of the inverse proximity effect.

\begin{table}
    \SetTblrInner{rowsep=0pt,colsep=0pt}
    \caption{All possible pair symmetries in the TMD-SC heterostructure. Correlations can be (E)ven or (O)dd, with the constraint that the product equals $-1$ according to Eq.~\eqref{eq:spot_rule}. \label{tab:pair_correlations_symmetries}}
    \begin{tblr}{colspec={X[l,m]X[c,m]X[c,m]X[c,m]X[c,m]}, width=\columnwidth}
        \hline \hline
                           & Spin    & Parity ($\mathbf{k}$) & Orbital & Time ($\omega$) \\ \hline
        {$F_{ab}^{(0)}$}   & Singlet & {E                                                \\ E \\ O \\ O} & {E \\ O \\ E \\ O}       & {E \\ O \\ O \\ E}            \\ \hline
        $F_{ab}^{(x,y,z)}$ & Triplet & {E                                                \\ E \\ O \\ O}                   & {E \\ O \\ E \\ O}       & {O \\ E \\ E \\ O}             \\ \hline \hline
    \end{tblr}
\end{table}

\subsection{Spectral signatures and density of states \label{sec:spectralA}}
The normal parts of the Green's function given by Eq.~\eqref{eq:GF} allows us to calculate the spectral function and density of states.
We calculate the spectral function of the total system as
\begin{equation}
    \label{eq:Akw}
    A(\mathbf{k}, \omega) = - \frac{1}{\pi} \text{Im}\, \text{Tr}\, G^R(\mathbf{k}, \omega),
\end{equation}
where $G^R(\mathbf{k}, \omega)$ is the retarded Green's function of the total system, obtained by taking the analytic continuation of the complex frequencies $z$ to the real axis, i.e, $z \to \omega + i \eta$ with a small damping factor $\eta > 0$ \cite{zagoskin,mahan2013many}.
Unless otherwise specified, we use $\eta = 1\,\mathrm{meV}$ and the exact result is obtained in the limit $\eta \to 0$.
The spectral function can be accessed by ARPES measurements \cite{hufner2013photoelectron,lv2019angle,yu2020relevance,RevModPhys.93.025006}.
For presentation purposes and to quantify the size of certain quantities, we define
\begin{equation}
    \label{eq:momentum_integral}
    \begin{split}
        \mathcal{X}(\omega) & = \frac{1}{\Omega_{\text{BZ}}} \int_{\text{BZ}} |X(\mathbf{k},\omega)| \, d^2k\,, \\
        \mathbb{X}          & =\int d\omega\,\mathcal{X}(\omega)\,.
    \end{split}
\end{equation}
Here, $\mathcal{X}(\omega)$ represents the Brillouin zone average of the absolute value of a momentum dependent quantity $X(\mathbf{k}, \omega)$, while $\mathbb{X}$ is its integration over frequency, and ${\Omega_{\text{BZ}} = 8\pi^2 / (\sqrt{3} a^2)}$ is the area of the Brillouin zone.
In an experimental setting, the density of states (DOS) is often more accessible than the spectral function, which can be obtained from the spectral function following Eq.~\eqref{eq:momentum_integral} via integration over the Brillouin zone
\begin{equation}
    D(\omega) \equiv \mathcal{A}(\omega) = \frac{1}{\Omega_{\text{BZ}}} \int_{\text{BZ}} A(\mathbf{k}, \omega) \, d^2k\,,
\end{equation}
where the absolute value is not needed since the spectral function is always positive.
The division by the area of the Brillouin zone ensures that we fulfill the normalization condition that the integral over frequency of the DOS is equal to the number of bands in the Hamiltonian, which is $16$ for our system when we take both spin and orbital degrees of freedom into account, as well as electron and hole bands in the Nambu basis.

In the following, we use the methodology discussed in this section to explore the impact of multiple orbitals, together with first Ising SOC and then also Rashba SOC, on the superconducting properties of the TMD-SC heterostructure system.


\begin{figure*}
    \centering
    \includegraphics{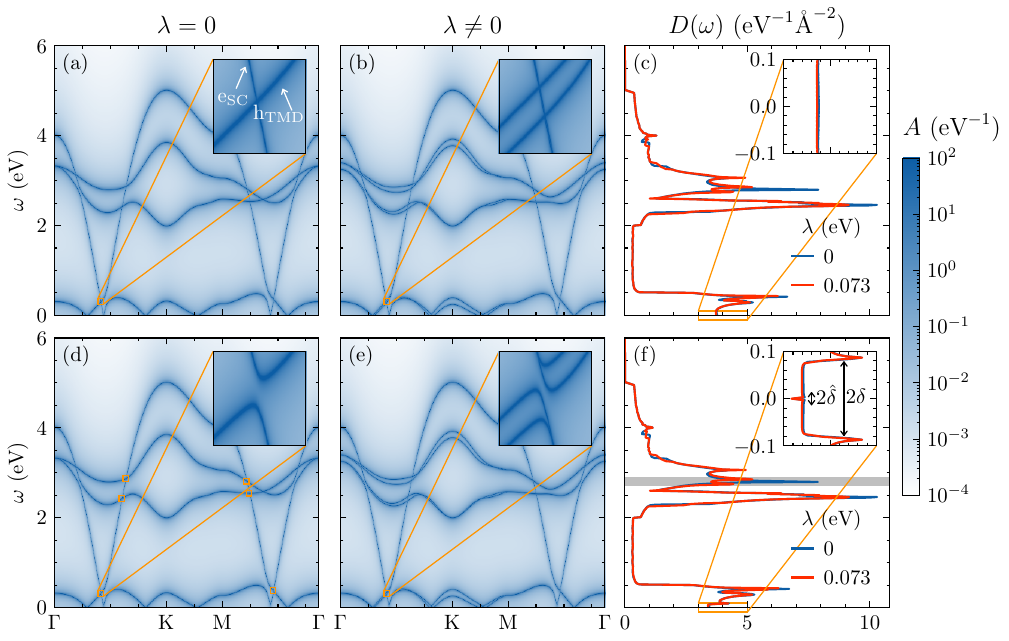}
    \caption{(a,b) Spectral function of the TMD-SC heterostructure as a function of frequency $\omega$ and momentum-path along high-symmetry lines for distinct values of the Ising SOC $\lambda$ without superconductivity $\Delta = 0$.
        Insets in (a,b) show a close-up of a crossing between electron ($e_{\text{SC}}$) and hole bands ($h_{\text{TMD}}$) in the SC and TMD.
        (c) DOS as a function of $\omega$ for distinct $\lambda$.
        Inset shows a detailed view of the DOS around zero frequency.
        (d-f) The same as in (a-c) for finite superconductivity $\Delta \neq 0$.
        Inset in (f) reveals the induced superconducting gaps $2\delta$ and $2\hat{\delta}$.
        Gray horizontal area indicates an energy window where Ising SOC results in spin splitting of peaks from Van Hove singularities.
        All plots are symmetric around zero frequency $\omega = 0$ due to particle-hole symmetry of $H_{\text{BdG}}$.
        The parameters used are as follows: $t = 0.2\,\text{eV}$, $\lambda = 0.073\,\text{eV}$, $a = 3.19\,\text{\AA}$, $\mu = -0.4\,\text{eV}$, $t_{\text{SC}} = 1\,\text{eV}$, $\mu_{\text{SC}} = 2\,\text{eV}$ and $\Delta = 0.1\,\text{eV}$ unless otherwise specified.
    }
    \label{fig:dos_combo}
\end{figure*}

\section{Effect of Ising SOC on spectral function and density of states \label{sec:results_ising}}
In order to understand how the multiorbital structure and Ising SOC interplay in the TMD-SC heterostructure, we start by inspecting the impact of the Ising SOC on the spectral function and DOS in the absence of any Rashba SOC, thus setting $\gamma = 0$ in Eq.~\eqref{eq:ham_rashba}.
Following Sec.~\ref{sec:spectralA}, we obtain the spectral function in Figs.~\ref{fig:dos_combo}(a,b,d,e) as a function of frequency $\omega$ and along a high-symmetry momentum-path for two distinct values of the Ising SOC strength $\lambda$.
The corresponding DOS as a function of $\omega$ is shown in Figs.~\ref{fig:dos_combo}(c,f) and the figure is organized such that we present in the first row the nonsuperconducting state $\Delta = 0$, while the lower row shows results for finite superconductivity $\Delta \neq 0$.
Note that the spectral functions and DOS are symmetric around zero frequency $\omega = 0$ due to particle-hole symmetry in the BdG formalism.

We start by analyzing the nonsuperconducting state $\Delta = 0$ in Fig.~\ref{fig:dos_combo}(a,b) where the three bands of the TMD and the band of the SC can easily be identified by a comparison with the individual band structures in Fig.~\ref{fig:model}(c).
The TMD bands are spin-split due to the Ising SOC for finite $\lambda \neq 0$, while the SC bands are spin degenerate, compare Figs.~\ref{fig:dos_combo}(a,b).
The spin splitting is particularly pronounced at the $K$ and $K'$ points and results from the TMD monolayer lacking inversion symmetry.
Due to the presence of time-reversal symmetry, the spin-$\uparrow$ state at $K$ corresponds to the spin-$\downarrow$ state at $K'$, which results in spin-valley locking.
Moreover, the TMD and SC bands exhibit crossings at both low and high frequencies, which are shifted in both frequency and momentum at nonzero Ising SOC, compare insets in Figs.~\ref{fig:dos_combo}(a,b).
Note that even with Ising SOC included, the spin-degeneracy is protected along $\overline{\Gamma M}$.
This is due to the combination of this line segment coinciding with a mirror plane of the underlying point group $D_{3\text{h}}$ of the monolayer and the Hamiltonian being diagonal in spin space when only Ising SOC is considered.
This results in both spin blocks being related by a unitary transformation and thus having the same eigenvalues, protecting the spin degeneracy along $\overline{\Gamma M}$.

Adding superconductivity with a finite $\Delta \neq 0$, we find a small finite gap around zero energy $\omega = 0$, which we analyze in more detail below when discussing the DOS.
The coupling between the TMD and the SC also results in hybridization gaps at finite energies, indicated with orange boxes in Fig.~\ref{fig:dos_combo}(d), one of which is shown in detail in the insets of Figs.~\ref{fig:dos_combo}(d,e).
In particular, these highlighted gaps occur at the band crossing between the electron SC and hole TMD bands, labeled as $e_{\text{SC}}$ and $h_{\text{TMD}}$ in Fig.~\ref{fig:dos_combo}(a).
Due to particle-hole symmetry, similar hybridization gaps occur at negative energies, with the difference that electron and hole characters of the bands are interchanged.
Without Ising SOC $\lambda = 0$ the hybridized bands are doubly degenerate, while without superconductivity $\Delta = 0$ no hybridization gap opens.
The hybridization occurs between one of the spin degenerate bands in the SC separately with one of the two TMD bands in the inset, a consequence of the Hamiltonian with Ising SOC being block-diagonal in spin space.
This results in the two highlighted hybridization gaps for $\lambda \neq 0$ in Fig.~\ref{fig:dos_combo}(e) not interacting with each other.
The Ising field has the effect that the hybridization gaps for the two different spin states are shifted both in energy and momentum with respect to each other.
This shift represents a signature of proximity-induced superconductivity in TMD monolayers that specifically relates to the appearance of spin-triplet correlations as we see later.

Further insights into the impact of the Ising field on the superconducting properties of proximitized TMD monolayers is obtained from the DOS $D(\omega)$, presented in Figs.~\ref{fig:dos_combo}(c,f).
The biggest influence of the Ising SOC is the lifting of the spin degeneracy of the TMD bands, which shifts the respective peaks resulting from Van Hove singularities in the band structure, highlighted by the gray horizontal area in Fig.~\ref{fig:dos_combo}(f).
In addition, comparing the insets of Figs.~\ref{fig:dos_combo}(c,f) shows that the superconducting gap is opened around zero frequency when $\Delta \neq 0$, as expected.
The size of this gap is $2 \delta \approx 0.174 \, \text{eV}$, which is smaller than the gap size for the isolated SC given by $2 \Delta = 0.2 \, \text{eV}$.
This is a result of the hybridization between the TMD and SC bands and not the usual proximity effect observed at the boundary of a superconducting junction, since our calculation only considers a single layer in the $z$ direction.
We find that the decrease in gap size can be understood by employing a minimal two-band model detailed in Appendix~\ref{sec:scgap}, which also reveals that the superconducting gap is shifted to larger momentum.
The inset in Fig.~\ref{fig:dos_combo}(f) shows also a second smaller gap $2 \hat{\delta}$, which is the induced superconducting gap in the TMD bands.
The size of this gap can be understood from the same minimal model in Appendix~\ref{sec:scgap} and is well approximated by the expression
\begin{equation}
    2 \hat{\delta} = 2 \frac{\Delta (t \cdot v)^2}{E_{\text{SC}}^2} \approx 1.5\,\text{meV}, \label{eq:induced_gap_size}
\end{equation}
where $E_{\text{SC}} = 2.07\,\text{eV}$ is the energy of the superconducting band at the specific momentum where the TMD conduction band crosses zero energy along $\overline{\Gamma M}$ and $v = 0.89$ is the weight of the TMD band on the $d_{z^2}$ orbital at this momentum.

Last, we note that there are no clear signatures of the hybridization gaps in the DOS.
This is explained by a detailed look at the spectral functions in Fig~\ref{fig:dos_combo}, which shows that the hybridization gap does not lie at isotropic values of frequency around the $\Gamma$ point.
Therefore, the hybridization gaps are smeared out in the DOS which integrates over the whole Brillouin zone, making them hard to observe in an experimental setting.
This indicates that similar hybridization gaps, that have been put forward to be an experimentally measurable signature of odd-frequency superconductivity on the basis of simpler models \cite{komendova_odd_frequency_2015, tang_magnetic_2021, PhysRevB.108.214510}, may be harder to observe than previously proposed.

\begin{figure}
    \centering
    \includegraphics[width=\columnwidth]{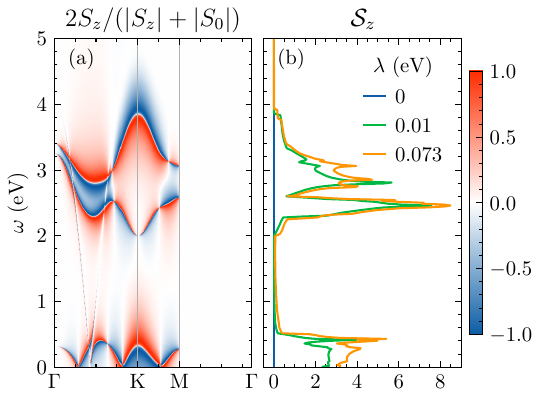}
    \caption{(a) Normalized spin density along $z$ of the TMD-SC heterostructure as a function of frequency and momenta along high-symmetry lines. (b) Absolute value of the spin density along $z$ integrated over the first Brillouin zone for different values of the Ising SOC strength $\lambda$. Parameters are the same as in Fig.~\ref{fig:dos_combo}.}
    \label{fig:spin_combo}
\end{figure}

Besides the spectral function and the DOS, we analyze the spin polarization of the heterostructure, by calculating the spin density
\begin{equation}
    S_j(\mathbf{k}, \omega) = -\frac{1}{\pi} \text{Im}\, \text{Tr}\,[G^R(\mathbf{k}, \omega) \sigma_{j}]
\end{equation}
with $S_{0}$ giving the spectral function in Eq.~\eqref{eq:Akw}.
Since the Ising SOC is along $z$, only the spin-triplet polarization along $z$ is finite, namely, $S_z(\mathbf{k}, \omega) \neq 0$.
In Fig.~\ref{fig:spin_combo}(a), we present the normalized spin density along $z$ as a function of $\omega$ and momenta along high-symmetry lines.
As seen, it is strongest where the bands are spin split, including for example around the $K$ points, signaling its origin due to the Ising SOC.
Since spin degeneracy is protected along $\overline{\Gamma M}$, the spin-$z$ density along this high-symmetry line is zero.

To further characterize the spin density along $z$, we calculate the absolute value of the spin density integrated over the first Brillouin zone following Eq.~\eqref{eq:momentum_integral} and denoted by $\mathcal{S}_{z}$. The absolute value is motivated by the fact that due to the presence of time-reversal symmetry, a integration without absolute value yields zero.
Fig.~\ref{fig:spin_combo}(b) shows $\mathcal{S}_{z}$ for distinct values of the Ising SOC $\lambda$.
Without Ising SOC $\lambda = 0$, there is zero spin-$z$ density and increasing the Ising SOC strength leads to a higher spin-polarization.
Since the integrated curves for the realistic value $\lambda = 0.073\,\text{eV}$ of the $\text{MoS}_2$ monolayer \cite{liu_three-band_2013} is not proportionally larger than a much smaller intermediate value, we conclude that the strength of Ising SOC in $\text{MoS}_2$ is sufficiently large to saturate how much spin polarization may be achieved in TMDs.


\begin{figure*}
    \centering
    \includegraphics[width=\textwidth]{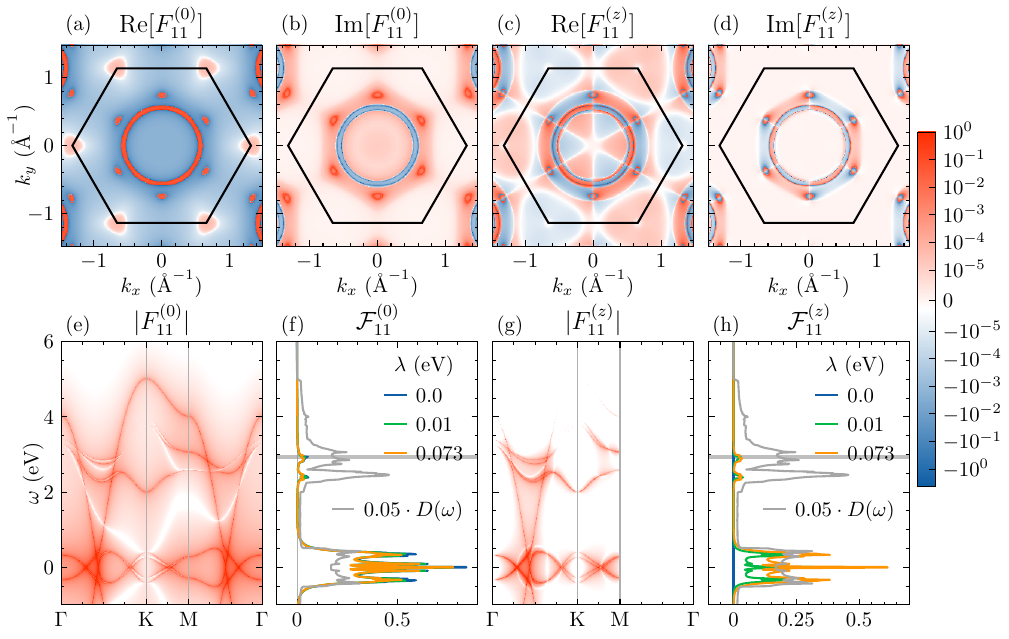}
    \caption{(a-d) Real and imaginary part of the spin-singlet $F^{(0)}_{11}$ and mixed spin-triplet $F^{(z)}_{11}$ pair amplitudes, respectively, in the $d_{z^2}$ orbital as a function of momentum at $\omega = 0.5\,\text{eV}$.
        Black hexagon indicates the first Brillouin zone.
        (e,g) Absolute value of the pair amplitudes $|F^{(0)}_{11}|$ and $|F^{(z)}_{11}|$, respectively, as a function of frequency and momenta along high-symmetry lines.
        (f,h) Pair amplitudes integrated over momentum $\mathcal{F}^{(0)}_{11}$ and $\mathcal{F}^{(z)}_{11}$, respectively, for different values of the Ising SOC strength $\lambda$ as well as the rescaled DOS $0.05 \cdot D(\omega)$ in gray for reference.
        Gray horizontal area indicates an energy window where hybridization gaps form that we analyze in further detail in Fig.~\ref{fig:paper1_EHNNUD11_pairings_abs}.
        All colored plots share the same color scaling. Parameters are the same as in Fig.~\ref{fig:dos_combo}.}
    \label{fig:spin_pol_combo}
\end{figure*}

\section{Effect of Ising SOC on induced pair amplitudes \label{sec:results_pair_correlations}}
Having established the presence of a finite spin polarization, we now explore how it relates to the emergence of triplet superconducting correlations.
Following the symmetry classification scheme introduced in Sec.~\ref{sec:methods_pair_correlations}, the anomalous correlations can be even or odd in all the four degrees of freedom spin, parity, orbital and frequency.
Tab.~\ref{tab:pair_correlations_symmetries} summarizes the allowed symmetry classes, which fulfill Fermi-Dirac statistics, as discussed in regards to Eq.~$\eqref{eq:spot_rule}$.
The combination of the four degrees of freedom allows for in total four spin-singlet and four spin-triplet pair symmetries.
While multiple classes of these also emerge in TMD-SC heterostructures, in order to understand the roles of the Ising field and multiple orbitals, we concentrate our analysis on the induced local superconducting correlations in the $d_{z^2}$ band ($a=b=1$) of the TMD.
Since this band is directly coupled to the superconductor via the tunnel amplitude $t$, it can reasonably be expected to host the strongest induced superconducting correlations in the TMD.
By looking at Tab.~\ref{tab:pair_correlations_symmetries} and keeping in mind that the pair amplitude $F_{11}^{(i)}$ is necessarily even in the orbital degree of freedom, the spin-singlet part of the correlations can be either even in momentum and frequency or odd in both, while the spin-triplet components are even in spin and therefore either momentum or frequency has to be odd while the other one is even.

To visualize the behavior of the induced pair amplitudes in the TMD, we show in Figs.~\ref{fig:spin_pol_combo}(a-d) the real and imaginary parts of the spin-singlet $F^{(0)}_{11}$ and mixed spin-triplet $F^{(z)}_{11}$ pair amplitudes as a function of momentum in the BZ at a prototypical value of frequency $\omega= 0.5\,\text{eV}$.
The equal in spin triplets $F^{(x,y)}_{11}$ are zero when only Ising SOC is present, as there is a $U(1)$-symmetry resulting from $[H,\sigma_z]=0$, forbidding equal spin-triplet pairing.
This also immediately implies that the spin polarization $P_{11}$ of the Cooper pair, as defined in Eq.~\eqref{eq:spin_pol}, has to be zero as well.
A closer inspection of the results shows that the spin-singlet (mixed spin-triplet) pair amplitudes are only even (odd) in momentum.
This is more constrained than the restrictions imposed by the SPOT classification as detailed in Tab.~\ref{tab:pair_correlations_symmetries}, which allows also for odd (even) in momentum spin-singlet (mixed spin-triplet) pair amplitudes, and therefore unveils that the pair amplitudes in Fig.~\ref{fig:spin_pol_combo} are specifically not odd in frequency.
We note here that this result is specific to the $F_{11}$ and also the $F_{44}$ pair amplitudes.
In contrast, the singlet even in orbital amplitudes $F_{22,33}^{(0)}$ are the sum of one part that is even in both momentum and frequency and one that is odd in both.
Furthermore, we verify that these numerical results also hold true via a direct analytical calculation, showing that there are additional constraints resulting from the structure of the Hamiltonian, which forbids odd-frequency pairing in certain cases.
An intriguing feature of the induced spin-singlet and spin-triplet pair amplitudes in the TMD is that they host direct information of the TMD Brillouin zone, with clear sign changes in the spin-triplet component when moving in momentum space, alike the normal state Berry curvature discussed in Ref.~\cite{liu_three-band_2013}.

We obtain further details of the pair amplitudes from Figs.~\ref{fig:spin_pol_combo}(e,g), where we show the absolute value of the spin-singlet and mixed spin-triplet amplitudes $|F^{(0,z)}_{11}|$ as a function of frequency $\omega$ along high-symmetry lines in the BZ.
In general, the induced superconducting correlations are the strongest closest to the hybridization gaps between TMD and SC and also near zero energy, which is directly tied to the proximity-induced superconductivity.
Because we consider a spin-singlet $s$-wave superconductor, we expect the induced singlet correlations in the TMD orbital $d_{z^2}$ to dominate over the triplet correlations.
While the absolute value of the mixed spin-triplet correlations are in general smaller than the induced singlet correlations, they are actually of comparable size close to the hybridization gaps.
In addition, no spin-triplet pairing is induced on the line $\overline{\Gamma M}$, see Fig.~\ref{fig:spin_pol_combo}(g), which is understood by the associated protection of the spin degeneracy, which was also reported in Ref.~\cite{PhysRevB.98.144518}. This highlights the connection between mixed spin-triplet correlations and spin-triplet density along $z$ as shown in Fig.~\ref{fig:spin_combo}.
In fact, contrasting Fig.~\ref{fig:spin_pol_combo} with the spin-triplet density in Fig.~\ref{fig:spin_combo}, we see that both finite spin polarization and how close the TMD and SC bands are in energy are essential factors in determining the strength of induced triplet correlations.

To further establish a connection with experimentally relevant quantities, we finally look at the momentum integrated spin-singlet and spin-triplet pair amplitudes $\mathcal{F}^{(0,z)}_{11}$ following Eq.~\eqref{eq:momentum_integral}.
Notably, the magnitude square of the pair amplitudes determines the probability of Andreev reflections through which it directly determines the current in NS-junctions or the Josephson current \cite{mahan2013many, PhysRevB.25.4515, PhysRevB.54.7366,PhysRevB.109.205406}, currents that are relevant for superconducting spintronics \cite{linder_superconducting_spintronics_2015,Eschrig2015,yang_superconducting_spintronics_2021,mel2022superconducting}.
Besides these experimentally relevant observables, the pair amplitude also directly affects response functions, such as in the Meissner effect and spin susceptibility, see Refs.~\cite{Cayao2020odd,triola_role_2020,linder_odd-frequency_2019} for an extensive discussion on the importance of the pair amplitude in experimental observables.
In Figs.~\ref{fig:spin_pol_combo}(f,h) we show $\mathcal{F}^{(0,z)}_{11}$ for different Ising SOC strengths $\lambda$ in addition to the DOS, which is rescaled to $0.05 \cdot D(\omega)$ to fit into the same plot for a comparison.
The spin-singlet pair amplitudes show far less dependence on the strength of the Ising SOC, while the mixed spin-triplet correlations vanish for $\lambda = 0$ and increase with $\lambda$, thus pointing to a connection between Ising SOC and mixed spin-triplet correlations that we explore in further detail below.
Peaks in the strength of the induced pair amplitudes proliferate at zero and finite frequencies where the hybridization gaps occur.
The thin area highlighted in gray in both Fig.~\ref{fig:spin_pol_combo}(f,h) indicates a frequency range where the spin-split TMD bands intersect with the superconducting bands and form hybridization gaps as indicated with orange boxes in Fig.~\ref{fig:dos_combo}(d).
Since the peaks in the DOS (from the Van Hove singularities) of the TMD bands lie just outside of this gray area, the increased superconducting correlations are not simply a result of a higher density of states but rather due to the hybridization between TMD and SC bands.

\begin{figure}
    \centering
    \includegraphics[width=\columnwidth]{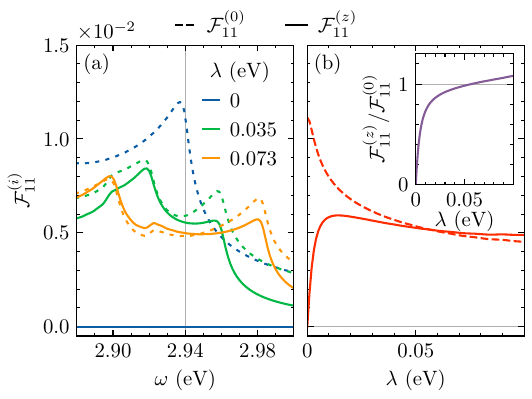}
    \caption{(a) Spin-singlet $\mathcal{F}^{(0)}_{11}$ and mixed spin-triplet $\mathcal{F}^{(z)}_{11}$ pair amplitudes in the $d_{z^2}$ orbital as a function of frequency $\omega$ for different values of the Ising SOC strength $\lambda$.
        (b) Pair amplitude strengths at fixed frequency $\omega = 2.94\,\text{eV}$ (vertical gray line in (a)) as a function of Ising SOC strength $\lambda$.
        The inset in (b) shows the ratio between mixed spin-triplets and singlet $\mathcal{F}^{(z)}_{11}/\mathcal{F}^{(0)}_{11}$. Parameters are the same as in Fig.~\ref{fig:dos_combo}.}
    \label{fig:paper1_EHNNUD11_pairings_abs}
\end{figure}

\subsection{Linking Ising SOC and mixed spin-triplet pair amplitudes \label{sec:linking_ising_triplet}}
In order to deepen the link between the spin splitting induced by Ising SOC and the emergence of mixed spin-triplet superconducting correlations, we show in Fig.~\ref{fig:paper1_EHNNUD11_pairings_abs}(a) the total spin-singlet and mixed spin-triplet pair amplitudes within the frequency range indicated by the horizontal shaded gray region in Figs.~\ref{fig:spin_pol_combo}(f,h).
Within this frequency window, the spin-split electron TMD bands intersect with the superconducting hole bands, forming a hybridization gap.
In the absence of Ising SOC, no mixed spin-triplet pairing forms (solid lines), as expected, and the spin-singlet pairing (dashed lines) form  a single peak.
By increasing $\lambda$, the spin degeneracy is lifted, which results in the splitting of the singlet peak into two peaks.
Interestingly, the position of the large singlet peak for $\lambda = 0$, as well as any of the obtained peaks, corresponds to the frequency values where the hybridization gaps between the TMD and the superconductor arise, see Fig.~\ref{fig:dos_combo}(d).

The spin-singlet pairing is in general dominant over the mixed spin-triplet component, but the triplet pairing can actually achieve larger values at frequencies between the two peaks.
To understand this, we look at the real part of $F^{\uparrow \downarrow}_{11}$ and $F^{\downarrow \uparrow}_{11}$, since the imaginary part only contributes at the hybridization gaps for $\eta \to 0^+$.
For frequencies far from the hybridization gaps, $F^{\uparrow \downarrow}_{11} \approx -F^{\downarrow \uparrow}_{11}$, which is exact at zero Ising SOC $\lambda = 0$.
As established above, the hybridization gaps occur at different frequencies $\omega_1 \neq \omega_2$ for the two spin states when $\lambda \neq 0$, see Fig.~\ref{fig:dos_combo}(e), where the separation of the values of $\omega_i$ results from the Ising SOC splitting the TMD bands and thus the separation between $\omega_1$ and $\omega_2$ increases with $\lambda$.
Due to the denominator of the Green's function, both pair correlations follow the general form $\propto (\omega - \omega_i)^{-1}$, leading to them diverging at different frequencies.
In both cases, the real parts of the pair correlations switches sign at $\omega_i$, implying that the real part of $F^{\uparrow \downarrow}_{11}$ and $F^{\downarrow \uparrow}_{11}$ between the two peaks have the same sign, one of them increasing with $\omega$ while the other one decreases.
As a consequence, the mixed spin-triplet pairing $F^{(z)}_{11} \propto F^{\uparrow\downarrow}_{11} + F^{\downarrow \uparrow}_{11}$ is finite in the whole interval between the two peaks, while the singlet component $F^{(0)}_{11} \propto F^{\uparrow\downarrow}_{11} - F^{\downarrow \uparrow}_{11}$ switches sign and therefore crosses zero at some in-between frequency.
This illustrates that it is not surprising to find dominating mixed spin-triplet correlations within the frequency interval between $\omega_1$ and $\omega_2$.
Importantly, since the dominating spin-triplet pairing is only due to the spin splitting of the hybridization gaps, we can directly link spin polarization along $z$ with the emergence of large mixed spin-triplet correlations. In addition, when measuring momentum-resolved superconducting correlations, missing a zero point between the two peaks due to spin split bands is evidence of the presence of mixed spin-triplet correlations.
It is however important to stress that the spin-singlet pair amplitude in Fig.~\ref{fig:paper1_EHNNUD11_pairings_abs}(a) does not vanish as we are extracting the superconducting correlations integrated over the full BZ, although it will be zero for specific momenta between the spin-split $\omega_i$.
The resulting finite value can again be attributed to the fact that the dispersion is not isotropic in momentum, which shifts both the values of $\omega_i$ and the frequency at which the singlet component is zero and then results in a finite value after momentum integration.

The role of the Ising SOC is further evidenced in Fig.~\ref{fig:paper1_EHNNUD11_pairings_abs}(b), where we plot the integrated spin-singlet and spin-triplet pair amplitudes as a function of the Ising SOC strength $\lambda$ at $\omega = 2.94\,\text{eV}$, roughly in between the two spin-split peaks in Fig.~\ref{fig:paper1_EHNNUD11_pairings_abs}(a), see vertical gray line.
The amount of spin-singlet correlations between the spin split peaks becomes weaker with increasing $\lambda$, while the mixed spin-triplet pairing becomes finite and even stronger than its singlet counterpart for a Ising SOC strength above $\lambda \approx 0.05\,\text{eV}$.
The inset shows the ratio of integrated mixed spin-triplet correlations to singlet correlations, which is greater than one for sufficiently large values of $\lambda$ and in particular also for the realistic value $\lambda = 0.073\,\text{eV}$ of the $\text{MoS}_2$ monolayer \cite{liu_three-band_2013}.
Therefore, mixed spin-triplet superconducting pair correlations represent an important part of the proximity-induced superconductivity in TMD monolayers. It may be identified by Andreev conductance \cite{PhysRevB.54.7366,PhysRevB.109.205406}.


\begin{figure}
    \centering
    \includegraphics[width=\columnwidth]{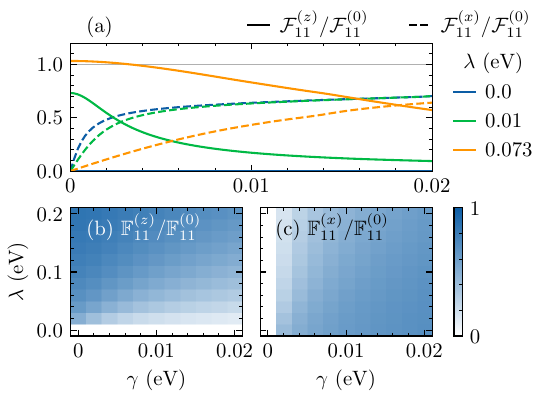}
    \caption{(a) Strength of mixed spin-triplet $\mathcal{F}_{11}^{(z)}$ and equal spin-triplet $\mathcal{F}_{11}^{(x)}$ amplitudes relative to the singlet amplitudes $\mathcal{F}_{11}^{(0)}$ as a function of Rashba SOC strength $\gamma$ at $\omega = 2.94\,\mathrm{eV}$.
        (b,c) Spin-triplet pair amplitudes as a function of both Ising SOC $\lambda$ and Rashba SOC $\gamma$ obtained by integrating over both momentum and frequency, normalized by the spin-singlet pair amplitude.
        Parameters are the same as in Fig.~\ref{fig:dos_combo}.}
    \label{fig:paper2_EHNNUD11_pairings_abs}
\end{figure}

\section{Effect of Rashba SOC \label{sec:results_rashba}}
In order to complete the picture of the emergence of spin-triplet superconducting correlations in the TMD-SC heterostructure, we next analyze the pair amplitudes in the presence of both Ising and Rashba SOC.
As explained in Sec.~\ref{sec:model}, the interface in the $z$ direction between TMD and SC, see Fig.~\ref{fig:model}(a), breaks the out-of-plane mirror symmetry with respect to the $xy$ plane, which induces Rashba SOC in the system.
Thus, the presence of Rashba SOC is an unavoidable effect in TMD-SC heterostructures.
Given its form in Eq.~\eqref{eq:ham_rashba}, Rashba SOC takes a manifestly off-diagonal form in spin space, expressed through the Pauli matrices $\sigma_x$ and $\sigma_y$.
This is in contrast to the Hamiltonian of the heterostructure with only Ising SOC included, which is diagonal in spin space.
Rashba SOC thus induces finite spin densities along $x$ and $y$, as it breaks the $U(1)$-symmetry resulting from $[H,\sigma_z]=0$ for the Hamiltonian of the system without Rashba SOC.
In addition, since the spin states being degenerate along $\overline{\Gamma M}$ requires the Hamiltonian to be diagonal in spin space, a finite Rashba SOC lifts this spin degeneracy, such that the $z$ spin density becomes nonzero also along this high-symmetry line.

From the connection between spin polarization and superconducting correlations established in Sec.~\ref{sec:linking_ising_triplet} as an effect of Ising SOC, it is natural to expect that Rashba SOC leads to equal spin-triplet pairing, as well as affected mixed spin-triplet pairing \cite{gorkov_superconducting_2001,reeg_proximity-induced_2015,PhysRevB.98.075425,triola_general_2016,Cayao2020odd,fksg-x8pr,Ahmed_2025}, which has already been discussed for intrinsic superconductivity in TMD monolayers in previous works \cite{PhysRevB.99.180505,PhysRevB.101.224503}.
Since both the $x$ and $y$ components of the spin Pauli matrices are included in the same manner in the Rashba SOC Hamiltonian, see Eq.~\eqref{eq:ham_rashba}, both equal spin-triplet pair amplitudes $F^{(x)}_{11}$ and $F^{(y)}_{11}$ are induced with the same strength, making it sufficient to focus on only one of them.

Figure~\ref{fig:paper2_EHNNUD11_pairings_abs}(a) shows the absolute value of the mixed $\mathcal{F}_{11}^{(z)}$ and equal spin-triplet $\mathcal{F}_{11}^{(x)}$ pair amplitudes integrated in momentum as a function of the Rashba SOC strength $\gamma$ for distinct values of the Ising SOC strength $\lambda$ at $\omega = 2.94 \, \text{eV}$.
In all cases, we normalize the spin-triplet components by the corresponding spin-singlet strength $\mathcal{F}_{11}^{(0)}$.
We observe that equal spin-triplets emerge in the system as soon as the Rashba SOC strength $\gamma$ is finite and that they undergo a fast increase at low $\gamma$ and then saturate, while the mixed spin-triplet pairing tends to slowly decrease with $\gamma$.
Note that at the crossing point, where the equal spin-triplet correlations become stronger than the mixed spin-triplet correlations, the value of Rashba SOC $\gamma$ is smaller than the respective value of Ising SOC $\lambda$.
Thus, even for relatively small values of Rashba SOC, as can be expected to be induced at the interface, equal spin-triplet correlations are significant.
Interestingly, the line with $\lambda = 0$ in Fig.~\ref{fig:paper2_EHNNUD11_pairings_abs}(a) even shows that without Ising SOC, Rashba SOC induces equal spin-triplet pairs but the mixed spin-triplet pairs are absent.
More details on the effect of Rashba SOC is presented in Appendix~\ref{sec:appendix_rashba}.

To further characterize the size of induced spin-triplet correlations in the TMD, we obtain the spin-triplet pair amplitudes integrated in momentum and frequency $\mathbb{F}_{11}^{(i)}$, according to Eq.~\eqref{eq:momentum_integral}.
In Figs.~\ref{fig:paper2_EHNNUD11_pairings_abs}(b,c) we show the spin-triplet pair amplitudes $\mathbb{F}_{11}^{(z,x)} / \mathbb{F}_{11}^{(0)}$, normalized by the spin-singlet pair amplitude.
With this, we corroborate the behavior of the spin-triplet amplitudes in Fig.~\ref{fig:paper2_EHNNUD11_pairings_abs}(a), showing that finite Rashba SOC induces only equal spin-triplet correlations, while Ising SOC induces only mixed spin-triplet correlations.
Furthermore, there is a trade-off between both types of triplet correlations, such that increasing one type suppresses the other one.
The relative strength of both types of spin-triplet correlations remains below one, which shows that the spin-singlet nature of the superconductor cannot be completely overcome.
However, using realistic values of both Ising and Rashba SOC, we find that both types of triplet correlations are still of comparable strength to the singlet correlations.
It is important to note here that the spin-triplet correlations being comparable to the spin-singlet ones is an observation specific to proximity-induced superconductivity in the TMD.
In contrast, the formation of spin-triplet pair correlations in the superconductor, as a result of the inverse proximity effect, is much smaller than the spin-singlet component; see Appendix~\ref{sec:appendix_sc} for a detailed discussion.


\section{Conclusions \label{sec:conclusion}}
Using a realistic three-orbital tight-binding model with anisotropic couplings to model a TMD monolayer, we investigate the superconducting properties that emerge when the TMD monolayer is proximitized with a spin-singlet $s$-wave superconductor.
In particular, we demonstrate that the multiorbital nature of the TMD monolayer leads to hybridization gaps that form at band crossings between electron (hole) superconducting and hole (electron) TMD bands at higher energies that are readily identified in the spectral function.
However, in contrast to previously considered simplified models, we find that the anisotropic couplings between bands within a more realistic model of the TMD monolayer make it difficult to identify the hybridization gaps in the density of states. Nonetheless, such hybridization gaps represent an exotic signature of proximity-induced superconductivity in the TMD monolayer.
Moreover, we analyze the effect of the characteristically large inherent Ising SOC of the TMD monolayer and show that it induces a spin splitting and spin polarization along $z$.
Having performed a symmetry classification of the induced superconducting correlations, we find that the presence of mixed spin-triplet pair amplitudes is directly linked to the presence of the finite spin polarization along $z$ and thus to the Ising SOC.
Using realistic values to model a monolayer of $\text{MoS}_2$, we show that the induced mixed spin-triplet correlations are of the same magnitude as their spin-singlet counterparts.
Furthermore, we demonstrate that these mixed spin-triplet correlations can even become larger between hybridization gaps coming from spin-split TMD bands, where the spin-singlet correlations instead cross zero.
This leads to a characteristic signature: the presence of mixed spin-triplet correlations can be inferred from the absence of a zero crossing of the singlet correlations between spin-split peaks.
Finally, since Rashba SOC may naturally emerge at the interface between the TMD monolayer and the superconductor, we study its effect on the induced superconducting correlations.
We show that it leads to the emergence of equal spin-triplet superconducting pairs. We further uncover a competing interplay between mixed and equal spin-triplet superconducting correlations, driven by the Ising and Rashba SOC, respectively.

We emphasize that the analysis performed here generalizes to other TMDs, such that our findings apply quite generally and suggest that proximitized monolayer TMDs are a promising platform for the realization of spin-triplet superconductivity as an alternative to the more commonly used superconductor-ferromagnet interfaces or TMDs with intrinsic superconductivity.
In contrast to intrinsic superconductivity, studying TMD monolayers in proximity to conventional superconductors permits us to understand the mechanism behind the emergence of spin-triplet pairs.
Also, our considered multiband model with anisotropic couplings between bands specifically allows us to assess higher energy gap structures as well as spin-triplet and spin-singlet pairs under realistic conditions, something that is beyond simplified models.
Therefore, our findings provide a deeper understanding of proximity-induced superconductivity in TMD monolayers, which can be relevant for assessing the interplay between strong SOC and superconductivity and also for superconducting spintronics \cite{budzin_proximity_2005,RevModPhys.77.1321,linder_superconducting_spintronics_2015,Eschrig2015,mel2022superconducting}.

\section*{\label{sec:data_availability} Data availability}
The data supporting this work, as well as the code to generate the figures, are openly available under \cite{kayatz_2026_18417351}.

\begin{acknowledgments}
    This work was partially supported by the Wallenberg Initiative Materials Science for Sustainability (WISE) funded by the Knut and Alice Wallenberg Foundation.
    J.C. acknowledges  financial support from the Swedish Research Council  (Vetenskapsr\aa det Grant No.~2021-04121) and from the Olle Engkvist Foundation (Grant No. 243-1026).
\end{acknowledgments}

\bibliography{library}

\newpage
\appendix

\section{\label{sec:tmdmodel} TMD model details}

In order to reproduce the low-energy band structure of the TMD monolayer, we use the three-band tight-binding model introduced in Ref.~\cite{liu_three-band_2013}, which includes the $d_{z^2}$, $d_{xy}$, and $d_{x^2-y^2}$ orbitals of the transition metal atoms.
\begin{figure}
    \centering
    \includegraphics[width=0.6\columnwidth]{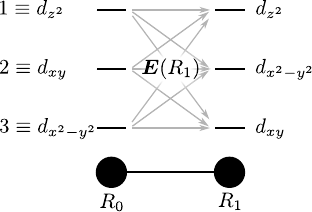}
    \caption{Sketch illustrating the nearest-neighbor hopping amplitudes between central site $R_0$ and nearest-neighbor site $R_1$, with the labeling of the TMD orbitals with indices $a=1,2,3$.}
    \label{fig:orbital_levels_2atoms}
\end{figure}
Fig.~\ref{fig:orbital_levels_2atoms} illustrates the hopping amplitudes between the orbitals on the central site $R_0$ and the neighboring atom at $R_1$, with the hopping matrix given by
\begin{equation}
    \mathbf{E}(R_1) = \begin{pmatrix}
        t_0  & t_1     & t_2    \\
        -t_1 & t_{11}  & t_{12} \\
        t_2  & -t_{12} & t_{22}
    \end{pmatrix}.
\end{equation}
Here the number of free parameters has already been constrained by the underlying symmetries. Hopping amplitudes to the other neighbors can be obtained from symmetry considerations as well, and  we label the second- and third-nearest neighbor hoppings as $r_{i}$ and $u_{i}$, respectively. Including on-site energies $\varepsilon_i$ and transforming to momentum space, the matrix elements of the Hamiltonian as given in Eq.~\eqref{eq:ham_tmd} are given by \cite{liu_three-band_2013}
{\allowdisplaybreaks
    \begin{align*}
        E^{11}_{\mathbf{k}} & = \varepsilon_1 + 2t_0 (\cos2\alpha + 2 \cos\alpha \cos\beta)                     \\
                            & + 2r_0 (\cos2\beta + 2 \cos3\alpha \cos\beta)                                     \\
                            & + 2u_0 (\cos4\alpha + 2 \cos2\alpha\cos2\beta)                                    \\
        E^{12}_{\mathbf{k}} & = 2it_1(\sin2\alpha + \sin\alpha\cos\beta) - 2\sqrt{3}t_2\sin\alpha\sin\beta      \\
                            & + 2 \sin3\alpha ((r_1+r_2) \sin\beta + i (r_1 - r_2)\cos\beta)                    \\
                            & + 2iu_1(\sin4\alpha + \sin2\alpha\cos2\beta) - 2\sqrt{3}u_2\sin2\alpha\sin2\beta  \\
        E^{13}_{\mathbf{k}} & = 2t_2(\cos2\alpha - \cos\alpha\cos\beta) + 2i \sqrt{3} t_1 \cos\alpha\sin\beta   \\
                            & - \frac{2}{\sqrt{3}} (r_1 + r_2) (\cos3\alpha \cos\beta - \cos2\beta)             \\
                            & + \frac{2}{\sqrt{3}} i(r_1 - r_2) \sin\beta (\cos3\alpha + 2\cos\beta)            \\
                            & + 2u_2(\cos4\alpha - \cos2\alpha\cos2\beta) + 2i\sqrt{3}u_1 \cos2\alpha\sin2\beta \\
        E^{22}_{\mathbf{k}} & = \varepsilon_2 + 2 t_{11} \cos2\alpha + (t_{11} +3t_{22}) \cos\alpha\cos\beta    \\
                            & + 4r_{11} \cos3\alpha\cos\beta + 2(r_{11} + \sqrt{3} r_{12})\cos2\beta            \\
                            & + 2u_{11} \cos4\alpha + (u_{11} +3u_{22}) \cos2\alpha\cos2\beta                   \\
        E^{23}_{\mathbf{k}} & = 4i t_{12} \sin\alpha(\cos\alpha - \cos\beta)                                    \\
                            & + \sqrt{3} (t_{22} - t_{11}) \sin\alpha\sin\beta                                  \\
                            & + 4r_{12}\sin3\alpha\sin\beta                                                     \\
                            & + 4iu_{12} \sin2\alpha (\cos2\alpha - \cos2\beta)                                 \\
                            & + \sqrt{3}(u_{22} - u_{11})\sin2\alpha\sin2\beta                                  \\
        E^{33}_{\mathbf{k}} & = \varepsilon_2 + 2 t_{22} \cos2\alpha + (3 t_{11} + t_{22}) \cos\alpha \cos\beta \\
                            & + 2r_{11}(\cos2\beta + 2 \cos3\alpha\cos\beta)                                    \\
                            & + \frac{2}{\sqrt{3}} r_{12} (4 \cos3\alpha \cos\beta - \cos2\beta)                \\
                            & + 2u_{22}\cos4\alpha + (3u_{11}+u_{22})\cos2\alpha\cos2\beta.
    \end{align*}
}
The remaining matrix elements are easily obtained through hermiticity $E^{ij}_{\mathbf{k}} = (E^{ji}_{\mathbf{k}})^*$. In these equations, the momentum dependence is expressed by $\alpha = k_x a/2$ and $\beta=\sqrt{3} k_y a/2$, as in the main text. We use the following values for the parameters \cite{liu_three-band_2013}
\begin{align*}
    t_0           & = -0.146 & r_0    & =0.06             & u_0    & =-0.038 \\
    t_1           & =-0.114  & r_1    & =-0.236           & u_1    & =0.046  \\
    t_2           & =0.506   & r_2    & =0.067            & u_2    & =0.001  \\
    t_{11}        & =0.085   & r_{11} & =0.016            & u_{11} & =0.266  \\
    t_{12}        & =0.162   & r_{12} & =0.087            & u_{12} & =-0.176 \\
    t_{22}        & =0.073   &        &                   & u_{22} & =-0.15  \\[2pt] \hline \\[-12pt]
    \varepsilon_1 & = 0.683  & a      & =3.19\,\text{\AA} &        &         \\
    \varepsilon_2 & = 1.707  & \mu    & =-0.4,            &        &
\end{align*}
which are all in units of $\mathrm{eV}$ if not specified otherwise. With the chosen values, the model reproduces the band structure of monolayer $\mathrm{MoS}_2$ as obtained from first principles \cite{liu_three-band_2013}. This allows us to use the model to characterize the superconducting correlations over the whole Brillouin zone.

\section{\label{sec:scgap} Minimal model for the sizes of the superconducting gaps}

As detailed in Sec.~\ref{sec:results_ising} in the main text, the hybridization between the TMD monolayer and the superconductor not only affects the size of the superconducting gap observed in the DOS but also results in an even smaller gap in the TMD conduction band at zero energy.
To understand both of these effects, we consider a minimal model that captures the essential physics of the hybridization between the TMD and the superconductor.

\subsection{Superconducting gap}

We start by explaining the change in size of the normal superconducting gap.
Since superconductivity is a low-energy phenomenon, we can approximate the band structure with its linear behavior close to the Fermi level, such that for the superconductor the electron and hole band together form a Dirac cone with Fermi velocity $v_{F}$, placed for simplicity at momentum $k=0$.
For the TMD we consider only the single low-energy band, which hybridizes the strongest with the superconducting band, approximated as a flat band at energy $-\varepsilon_{\text{TMD}}$ at the same momentum $k=0$. With this, we can write the Hamiltonian of the minimal model as:
\begin{equation}
    H_{\text{min}} = \begin{pmatrix}
        -\varepsilon_{\text{TMD}} & 0                        & t        & 0        \\
        0                         & \varepsilon_{\text{TMD}} & 0        & -t       \\
        t                         & 0                        & -v_{F} k & \Delta   \\
        0                         & -t                       & \Delta   & +v_{F} k \\
    \end{pmatrix}, \label{eq:app_ham_minimal_model}
\end{equation}
with $t$ controlling the coupling strength between the TMD and the superconductor, $\Delta$ being the superconducting order parameter, and the electron and hole bands ordered as ${\{e_{\text{TMD}},h_{\text{TMD}},e_{\text{SC}},h_{\text{SC}}\}}$ in the basis. From this, it is straightforward to calculate the superconducting gap as a difference of eigenvalues, resulting in
\begin{align}
    \delta & = \sqrt{2} \sqrt{\chi - \sqrt{\chi^2 - 4 \nu}},                          \\
    \chi   & = \varepsilon_{\text{TMD}}^2 + 2 t^2 + \Delta^2 + k^2 v_{F}^2, \nonumber \\
    \nu    & = (t^4 - 2 \varepsilon_{\text{TMD}} k t^2 v_{F} +
    \varepsilon_{\text{TMD}}^2 (\Delta^2 + k^2 v_{F}^2)). \nonumber
\end{align}
Analyzing this expression, we find that for zero coupling $t=0$ we recover the prototypical BCS gap of the form $2 \sqrt{\Delta^2 + (k v_{F})^2}$.
In the case of finite coupling we find that $\mathrm{d} \delta / \mathrm{d}k < 0$, which has two important consequences.
First, the position of the gap is moved to larger values of momentum $k$ and, second, the size of the gap is decreased compared to the uncoupled case.
This is because, of the two band crossings of the electron TMD band at $-\varepsilon_{\text{TMD}}$ with the Dirac cone, the one at $k>0$ is mediated directly through the coupling $t$, since here it crosses the electron band of the superconductor.
On the other hand, at $k<0$ this band crosses the hole band of the superconductor, which requires a higher-order process involving the superconducting pairing $\Delta$ to open the avoided level crossing.
As such, the hybridization gap opening is larger for the band crossing at $k>0$, which pushes the upper band further up, resulting in the shift of the minimum superconducting gap to larger values of $k$ and a decrease in gap size.

\subsection{Induced gap}

Trough the coupling to the superconductor, an additional smaller gap is induced in the TMD conduction band at zero energy, as shown in Fig.~\ref{fig:dos_combo}(f) in the main text.
To explain the size of this gap, we can again use the minimal model given in Eq.~\eqref{eq:app_ham_minimal_model}, where we now focus on the regime where $\epsilon_{\text{TMD}}$ is small compared to $v_F k$. In order to find a good numerical approximation of the induced gap size, we also include the modification $t \to t \cdot v$ where $v$ is the weight of the conduction band on the $d_{z^2}$ orbital (the one directly coupled to the superconductor).
We perform a Schrieffer-Wolff transformation up to second order in $t \cdot v$ to obtain an effective Hamiltonian for the TMD conduction band.
From this, we can approximate the size of the induced gap as:
\begin{equation}
    2\hat{\delta} = 2 \frac{\Delta (t \cdot v)^2}{(k v_{F})^2}.
\end{equation}
Taking the energy of the superconducting band to be constant $k v_{F} \approx E_{\text{SC}}$, we arrive at Eq.~\eqref{eq:induced_gap_size} in the main text.

\section{\label{sec:appendix_t}Effect of tunneling strength between TMD and SC}

\begin{figure}
    \centering
    \includegraphics[width=\columnwidth]{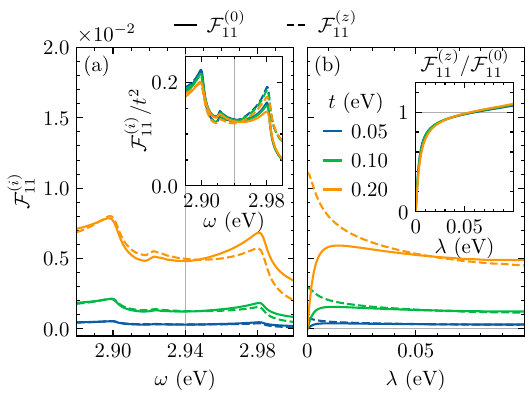}
    \caption{(a) Spin-singlet $\mathcal{F}_{11}^{(0)}$ and mixed spin-triplet $\mathcal{F}_{11}^{(z)}$ pair amplitudes in the $d_{z^2}$ orbital as a function of frequency $\omega$ for different values of the tunneling strength $t$. Inset shows the pair amplitudes divided by $t^2$.
        (b) Pair amplitude strengths at fixed frequency $\omega = 2.94\,\text{eV}$ (vertical gray line in (a)) as a function of Ising SOC strength $\lambda$.
        Inset in (b) shows the ratio between mixed spin-triplet and singlet $\mathcal{F}_{11}^{(z)}/\mathcal{F}_{11}^{(0)}$.
        Parameters are the same as in Fig.~\ref{fig:dos_combo} and this figure is also similar to Fig.~\ref{fig:paper1_EHNNUD11_pairings_abs}, but here we instead plot the pair amplitudes for different values of tunneling strength $t$.}
    \label{fig:vary_t}
\end{figure}
As specified in Sec.~\ref{sec:model}, we have chosen the value $t=0.2\,\text{eV}$ for the tunneling strength between the TMD and the superconductor in the numerical calculations reported in the main text.
This value is similar in size to the other hopping amplitudes in the TMD, see Appendix~\ref{sec:tmdmodel}, and a reasonable assumption albeit at the upper limit of what can be considered realistic in an experiment. In this appendix, we elaborate on the effect of varying the tunneling strength $t$ on the main results of the paper.

First, the discussion of the induced superconducting gap in Appendix~\ref{sec:scgap} leads to the approximation given by Eq.~\eqref{eq:induced_gap_size}, with the size of the induced gap being proportional to $t^2$.
This can intuitively be understood by noting that the induced superconducting correlations in the TMD are a second-order effect because the TMD is not intrinsically superconducting, so the tunneling strength $t$ enters twice in the process of forming a Cooper pair in the TMD.
With this, pairing terms can be seen as a second-order perturbation for small values of $t$.
In addition to the size of the induced gap, the same reasoning is valid for the size of the induced pair amplitudes in the TMD, which has direct consequence for the discussion of their magnitude.
Fig.~\ref{fig:vary_t}(a) confirms this by showing the spin-singlet $\mathcal{F}_{11}^{(0)}$ and mixed spin-triplet $\mathcal{F}_{11}^{(z)}$ pair amplitudes in the $d_{z^2}$ orbital as a function of frequency $\omega$ for different values of the tunneling strength $t$, which clearly show that the size of the pair amplitudes depends on $t$.
The inset in Fig.~\ref{fig:vary_t}(a) shows the pair amplitudes divided by $t^2$, which confirms that the size of the pair amplitudes is indeed well approximated by a $t^2$-dependence.
Further, Fig.~\ref{fig:vary_t}(b) shows the pair amplitude strengths at fixed frequency $\omega = 2.94\,\text{eV}$ as a function of Ising SOC strength $\lambda$, which also shows a dependence on $t$.
Since the inset in Fig.~\ref{fig:vary_t}(b) shows the ratio between mixed spin-triplet and singlet $\mathcal{F}_{11}^{(z)}/\mathcal{F}_{11}^{(0)}$, the result becomes largely independent of $t$.
This shows that wherever we calculate the absolute size of the pair amplitudes, see Figs.~\ref{fig:spin_pol_combo} and~\ref{fig:paper1_EHNNUD11_pairings_abs}, the resulting values are in good approximation proportional to $t^2$, while, wherever we calculate the relative size of the pair amplitudes, see Figs.~\ref{fig:paper1_EHNNUD11_pairings_abs}(b) and inset in~\ref{fig:paper2_EHNNUD11_pairings_abs}, the dependence of the results on $t$ is negligible.
In particular, this establishes that the comparison of the size of the induced pair amplitudes in the TMD is independent of the choice of $t$.
Since we have demonstrated a clear understanding of the effect of $t$ on the induced pair amplitude, this justifies our choice of $t=0.2\,\text{eV}$ in the main text, where a smaller value would only increase the computational cost without providing any additional physical understanding.

\section{\label{sec:appendix_rashba} Equal spin-triplet correlations with Rashba SOC}

\begin{figure*}
    \centering
    \includegraphics[width=\textwidth]{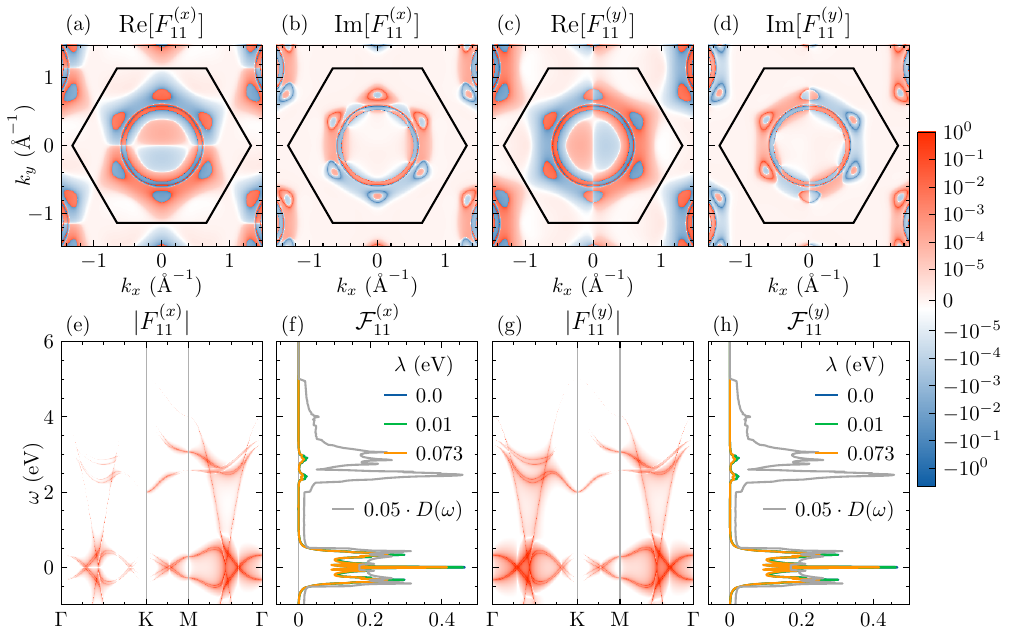}
    \caption{
        (a-d) Real and imaginary part of the equal spin-triplet $F^{(x,y)}_{11}$ pair amplitudes in the $d_{z^2}$ orbital as a function of momentum at $\omega = 0.5\,\text{eV}$.
        Including Rashba SOC allows the correlations to be odd in frequency, and therefore the correlations are in contrast to Fig.~\ref{fig:spin_pol_combo} superpositions of both even and odd in momentum contributions.
        Black hexagon indicates the first Brillouin zone.
        (e,g) Absolute value of the pair amplitudes $|F_{11}^{(x,y)}|$ as a function of frequency and momenta along high-symmetry lines.
        (f,h) Pair amplitudes integrated over momentum $\mathcal{F}^{(x,y)}_{11}$ for different values of the Ising SOC strength $\lambda$ as well as the rescaled DOS $0.05 \cdot D(\omega)$ in gray for reference. All colored plots share the same color scaling. Parameters are the same as in Fig~\ref{fig:dos_combo}.
        Similar to Fig.~\ref{fig:spin_pol_combo}, but here plotting equal spin-triplet pair amplitudes $F^{(x,y)}_{11}$ for finite Rashba SOC strength $\gamma = 0.01\,\text{eV}$ instead of singlet $F^{(0)}_{11}$ and mixed spin-triplet $F^{(z)}_{11}$ pair amplitudes in Fig.~\ref{fig:spin_pol_combo}.}
    \label{fig:spin_pol_combo2}
\end{figure*}

Here we present additional plots that show more details of the effect of Rashba SOC on the superconducting correlations, as discussed in Sec.~\ref{sec:results_rashba}.
In Fig.~\ref{fig:spin_pol_combo2} we plot the equal spin-triplet pair amplitudes $F^{(x,y)}_{11}$ in the $d_{z^2}$ orbital for finite Rashba SOC strength $\gamma = 0.01\,\text{eV}$.
This figure can be directly compared to Fig.~\ref{fig:spin_pol_combo} in the main text, where we instead plot $F^{(0,z)}_{11}$ for $\gamma = 0$.
Both figures share similar features, with some important differences.
First, in Sec.~\ref{sec:results_pair_correlations} we established that both $F^{(0,z)}_{11}$ are specifically not odd in frequency due to the structure of the Hamiltonian with only Ising SOC included.
With Rashba SOC included, this restriction is lifted such that these pair amplitudes can also be odd in frequency.
Thus $F^{(x,y)}_{11}$ are superpositions of both even and odd in momentum contributions.
Second, including Rashba SOC lifts the spin degeneracy along $\overline{\Gamma M}$, which allows for finite equal spin-triplet correlations along this line as well.

\section{\label{sec:appendix_sc}Anomalous correlations in the superconductor}

\begin{figure}
    \centering
    \includegraphics[width=\columnwidth]{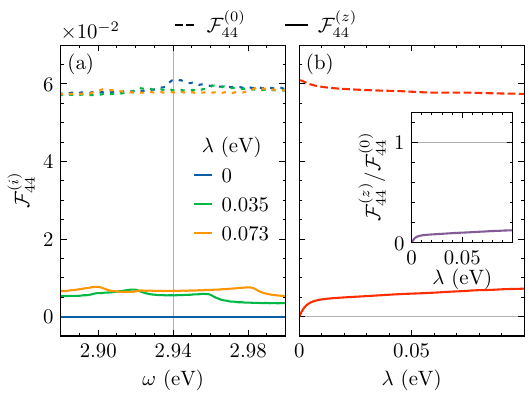}
    \caption{(a) Spin-singlet $\mathcal{F}_{44}^{(0)}$ and mixed spin-triplet $\mathcal{F}_{44}^{(z)}$ pair amplitudes in the superconducting band as a function of frequency $\omega$ for different values of the Ising SOC strength $\lambda$.
        (b) Pair amplitude strengths at fixed frequency $\omega = 2.94\,\text{eV}$ (vertical gray line in (a)) as a function of Ising SOC strength $\lambda$.
        The inset in (b) shows the ratio between mixed spin-triplets and singlet $\mathcal{F}_{44}^{(z)}/\mathcal{F}_{44}^{(0)}$.
        Parameters are the same as in Fig.~\ref{fig:dos_combo} and this figure is similar to Fig.~\ref{fig:paper1_EHNNUD11_pairings_abs}, but here we plot the pair amplitudes in the superconducting band instead of the $d_{z^2}$ orbital.
    }
    \label{fig:paper1_EHSSUD44_pairings_abs}
\end{figure}
As mentioned in Sec.~\ref{sec:results_rashba}, spin-triplet pair amplitudes form also in the superconductor as a result of the inverse proximity effect.
We illustrate this first in Fig.~\ref{fig:paper1_EHSSUD44_pairings_abs}, which shows that Ising SOC in the TMD also induces mixed spin-triplet pair amplitudes in the SC bands through the inverse proximity effect.
This figure can directly be compared to Fig.~\ref{fig:paper1_EHNNUD11_pairings_abs} in the main text, where we instead plot the pair amplitudes in the $d_{z^2}$ orbital of the TMD.
In comparison to the pair amplitudes for the $d_{z^2}$ orbital, the singlet correlations stay dominant in the superconductor, with the magnitude of the mixed spin-triplet correlations only approximately at 10\% of the strength of the singlet correlations at realistic values of the Ising SOC strength $\lambda$.
\begin{figure}
    \centering
    \includegraphics[width=\columnwidth]{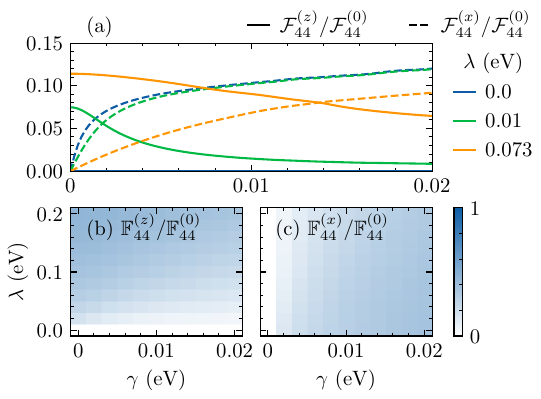}
    \caption{(a) Strength of the mixed spin-triplet $\mathcal{F}_{44}^{(z)}$ and equal spin-triplet $\mathcal{F}_{44}^{(x)}$ amplitudes relative to the singlet amplitudes $\mathcal{F}_{44}^{(0)}$ as a function of Rashba SOC strength $\gamma$ at $\omega = 2.94 \, \mathrm{eV}$.
        (b,c) Spin-triplet pair amplitudes as a function of both Ising SOC $\lambda$ and Rashba SOC $\gamma$ obtained by integrating over both momentum and frequency, normalized by the spin-singlet pair amplitude.
        Remaining parameters are the same as in Fig.~\ref{fig:dos_combo} and this figure is similar to Fig.~\ref{fig:paper2_EHNNUD11_pairings_abs}, but here we plot the pair amplitudes in the superconducting band instead of the $d_{z^2}$ orbital.
    }
    \label{fig:paper2_EHSSUD44_pairings_abs}
\end{figure}
Second, in Fig.~\ref{fig:paper2_EHSSUD44_pairings_abs} we analyze the effect of Rashba SOC, which can be directly compared to Fig.~\ref{fig:paper2_EHNNUD11_pairings_abs} in the main text.
We find that this figure shows qualitatively the same behavior as seen in Fig.~\ref{fig:paper2_EHNNUD11_pairings_abs}. Rashba SOC induces equal spin-triplet correlations but their magnitude stays similar in size to the mixed spin-triplet correlations.

\end{document}